\documentclass[prl,twocolumn,preprintnumbers,floatfix]{revtex4}

\usepackage[hyperfootnotes=false]{hyperref}
\usepackage{amsmath}
\usepackage{graphicx}

\newcommand{\eq}[1]{Eq.~\eqref{eq:#1}}
\newcommand{\eqs}[2]{Eqs.~\eqref{eq:#1} and \eqref{eq:#2}}
\newcommand{\fig}[1]{Fig.~\ref{fig:#1}}
\newcommand{\figs}[2]{Figs.~\ref{fig:#1} and \ref{fig:#2}}

\newcommand{\abs}[1]{\lvert#1\rvert}
\newcommand{\ord}[1]{\mathcal{O}(#1)}

\newcommand{\df}{\mathrm{d}}
\newcommand{\img}{\mathrm{i}}

\newcommand{\cI}{{\mathcal I}}

\newcommand{\GeV}{\,\mathrm{GeV}}
\newcommand{\TeV}{\,\mathrm{TeV}}

\newcommand{\nn}{\nonumber}

\newcommand{\lqcd}{\Lambda_\mathrm{QCD}}

\newcommand{\Ecm}{E_\mathrm{cm}}
\newcommand{\cut}{\mathrm{cut}}

\begin{document}


\preprint{\vbox{\hbox{MIT--CTP 4123}\hbox{May 21, 2010}}}

\title{The Beam Thrust Cross Section for Drell-Yan at NNLL Order}

\author{Iain W.~Stewart}
\affiliation{Center for Theoretical Physics, Massachusetts Institute of
  Technology, Cambridge, Massachusetts 02139, USA\vspace{-0.5ex}}

\author{Frank J.~Tackmann}
\affiliation{Center for Theoretical Physics, Massachusetts Institute of
  Technology, Cambridge, Massachusetts 02139, USA\vspace{-0.5ex}}

\author{Wouter J.~Waalewijn\vspace{0.5ex}}
\affiliation{Center for Theoretical Physics, Massachusetts Institute of
  Technology, Cambridge, Massachusetts 02139, USA\vspace{-0.5ex}}

\begin{abstract}

At the LHC and Tevatron strong initial-state radiation (ISR) plays an
important role. It can significantly affect the partonic luminosity available
to the hard interaction or contaminate a signal with additional jets and soft
radiation.  An ideal process to study ISR is isolated Drell-Yan production, $pp\to
X\ell^+\ell^-$ without central jets, where the jet veto is provided by the
hadronic event shape beam thrust $\tau_B$. Most hadron collider event shapes
are designed to study central jets. In contrast, requiring
$\tau_B\ll 1$ provides an inclusive veto of central jets
and measures the spectrum of ISR. For $\tau_B\ll 1$ we carry out a resummation
of $\alpha_s^n\ln^m\!\tau_B$ corrections at
next-to-next-to-leading-logarithmic order. This is the first
resummation at this order for a hadron-hadron collider event shape.
Measurements of $\tau_B$ at the Tevatron and LHC can provide crucial tests of
our understanding of ISR and of $\tau_B$'s utility as a central jet veto.

\end{abstract}

\maketitle

\paragraph*{Introduction.}

Event shapes play a vital role in the success of QCD measurements at $e^+e^-$
colliders.  This includes the measurements of $\alpha_s(m_Z)$, the QCD
$\beta$ function and color factors~\cite{alphas},
and the tuning and testing of Monte Carlo event generators (see e.g.\ Refs.~\cite{tuning}).
Event shapes for the
more complicated environment at hadron\linebreak colliders have been designed and studied
in Refs.~\cite{Nagy:2003tz, Banfi:2004nk, Dissertori:2008es, Stewart:2009yx}.
There is much anticipation that they can play a significant
role at the Tevatron and LHC by improving our understanding of basic aspects of
QCD in high-energy collisions such as the underlying event and initial- and
final-state radiation, as well as nonperturbative effects. Here we focus on
initial-state radiation (ISR).  Strong ISR can significantly alter the partonic
luminosity available for the hard interaction. Additional jets from ISR can also
contaminate the jet signature for a specific signal.
An ideal process to study ISR is isolated Drell-Yan production, $pp\to X\ell^+\ell^-$ with
a veto on central jets.  By vetoing hard central jets, the measurement becomes
directly sensitive to how energetic and soft ISR contributes to the hadronic
final state $X$.

\begin{figure}[b!]
\includegraphics[width=0.8\columnwidth]{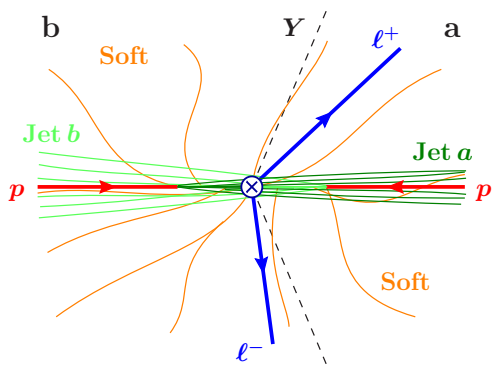}%
\vspace{-0.5ex}
\caption{\label{fig:isoDY} Isolated Drell-Yan production with a veto on central jets.}
\end{figure}

Recently an inclusive hadron collider event shape $\tau_B$
was introduced, called ``beam thrust''~\cite{Stewart:2009yx}.  For $\tau_B\ll 1$, the hadronic final
state consists of two back-to-back jets centered around the beam axis. The ISR
causing these jets occurs at measurable rapidities. In this limit $\tau_B$ provides
similar information as thrust in $e^+e^-\to \text{jets}$ with the thrust axis fixed to the
proton beam axis. For $\tau_B\sim 1$, the final state has energetic jets at
central rapidities.  Thus, requiring small $\tau_B$ provides an inclusive veto
on central jets, while allowing ISR in the forward direction, as depicted in
\fig{isoDY}.

Experimentally, beam thrust is one of the simplest hadronic observables at a
hadron collider. It requires no jet algorithms, is boost invariant along the
beam axis, and can be directly compared to theory predictions that require
no additional parton showering or hadronization from Monte Carlo programs.
Beam thrust is defined as~\cite{Stewart:2009yx}
\begin{equation} \label{eq:tauB}
\tau_B = \frac{1}{Q} \sum_k \abs{\vec{p}_{kT}} e^{-\abs{\eta_k - Y}}
\,,\end{equation}
where $Q^2$ and $Y$ are the dilepton invariant mass and rapidity, respectively.
The sum runs
over all (pseudo)particles in the final state except the two signal leptons,
where $\abs{\vec{p}_{kT}}$ and $\eta_k$ are the measured transverse momenta and
rapidities with respect to the beam axis, and all particles are considered
massless. The absolute value in the exponent in \eq{tauB} effectively divides
all particles into two hemispheres $\eta_k > Y$ and $\eta_k < Y$, where the
former gives $\abs{\vec{p}_{kT}} e^{-\eta_k} = E_k - p_k^z$ and the latter
$\abs{\vec{p}_{kT}} e^{\eta_k} = E_k + p_k^z$:
\begin{equation} \label{eq:tauB_alt}
\tau_B = \frac{1}{Q} \biggl[e^Y \!\sum_{\eta_k > Y} \!(E_k - p_k^z)
+ e^{-Y} \!\sum_{\eta_k < Y}\! (E_k + p_k^z) \biggr]
\,.\end{equation}
The dependence on $Y$ explicitly takes into account the boost of the partonic
center-of-mass frame, i.e.\ the fact that the collinear ISR in the direction of
the boost is narrower, as depicted in \fig{isoDY}.  From \eqs{tauB}{tauB_alt} we
see that soft particles with energies $E_k \ll Q$ as well as energetic particles
in the forward directions with $E_k - \abs{p_k^z} \ll Q$ contribute only small
amounts to $\tau_B$.  In particular, unmeasured particles beyond the rapidity
reach of the detector are exponentially suppressed, $\abs{\vec{p}_{kT}}
e^{-\abs{\eta_k}} \approx 2E_k e^{-2\abs{\eta_k}}$, and give negligible
contributions to $\tau_B$. On the other hand, energetic particles in the central
region with $E_k \pm p_k^z \sim E_k \sim Q$ give an $\ord{1}$ contribution to
$\tau_B$. Hence, a cut $\tau_B \leq \tau_B^\cut \ll 1$ vetoes central energetic
jets without requiring a jet algorithm.

Beam thrust is also theoretically clean. It is infrared safe, and an all-orders
factorization theorem exists for the cross section at small
$\tau_B$~\cite{Stewart:2009yx}. This allows for a higher-order summation of
large logarithms, $\alpha_s^n\ln^m\!\tau_B$, and the
calculation of perturbative and estimation of nonperturbative contributions
from soft radiation. The state of the art for resummation in hadron collider event shapes is the
next-to-leading logarithm (NLL) plus next-to-leading order (NLO)
analysis in Ref.~\cite{Banfi:2004nk}. In this Letter we present results
for the beam thrust cross section for $\tau_B\ll 1$ at
next-to-next-to-leading-logarithmic (NNLL) order. This represents the first complete calculation
to this order for a hadron collider event shape.  Letting $v_B-\img 0$ be the
Fourier conjugate variable to $\tau_B$, the Fourier-transformed cross section
exponentiates and has the form
\begin{equation}
\ln\frac{\df\sigma}{\df v_B} \sim L (\alpha_s L)^k + (\alpha_s L)^k + \alpha_s (\alpha_s L)^k + \dotsb
\,,\end{equation}
where $L = \ln v_B$ and we sum over $k \geq 1$. Here, the three sets of terms
are the leading logarithmic (LL), NLL, and NNLL corrections.

\paragraph*{Beam Thrust Factorization Theorem.}

The Drell-Yan beam thrust cross section for small $\tau_B$ obeys
the factorization theorem~\cite{Stewart:2009yx}
\begin{align} \label{eq:DYbeamrun}
\frac{\df\sigma}{\df Q \df Y \df \tau_B}
&= \frac{8\pi\alpha_\mathrm{em}^2}{9\Ecm^2 Q}
\sum_{ij} H_{ij}(Q^2, \mu_H)\, U_H(Q^2, \mu_H, \mu)
\nn\\ &\quad \times
\int\!\df t_a \df t_a'\, B_i(t_a - t_a', x_a, \mu_B)\, U^i_B(t_a', \mu_B, \mu)
\nn\\ &\quad \times
\int\!\df t_b \df t_b'\, B_j(t_b - t_b', x_b, \mu_B)\, U^j_B(t_b', \mu_B, \mu)
\nn\\ &\quad \times
\int\!\df k\,
Q S_B\Bigl(\tau_B Q - \frac{t_a + t_b}{Q} - k, \mu_S \Bigr)
\nn\\ &\quad \times
U_S(k, \mu_S, \mu)
\,,\end{align}
where $x_a = (Q/\Ecm) e^{Y}$ and $x_b = (Q/\Ecm) e^{-Y}$,
$\Ecm$ is the total center-of-mass energy, and the sum runs over quark flavors
$ij = \{u\bar u, \bar u u, d\bar d, \ldots\}$.  The hard function $H_{ij}(Q^2,
\mu_H)$ contains virtual radiation at the hard scale $Q$ (and also includes the
leptonic process).

The beam functions $B_i(t_a, x_a, \mu_B)$ and $B_j(t_b, x_b, \mu_B)$ in
\eq{DYbeamrun} depend on the momentum fractions $x_{a,b}$ and virtualities
$t_{a,b}$ of the partons $i$ and $j$ annihilated in the hard interaction. They
can be calculated in an operator-product expansion~\cite{Fleming:2006cd, Stewart:2010qs}
\begin{equation} \label{eq:B_fact}
B_i(t_a, x_a, \mu_B)
= \sum_k\!\int_{x_a}^1 \frac{\df\xi_a}{\xi_a}\,
 \cI_{ik}\Bigl(t_a,\frac{x_a}{\xi_a},\mu_B \Bigr) f_k(\xi_a, \mu_B)
\,,\end{equation}
and analogously for $B_j$. Here, the sum runs over parton species $k = \{g, u,
\bar u, d, \bar d, \ldots \}$ and $f_k(\xi_a, \mu_B)$ denotes the standard
parton distribution function (PDF) for parton $k$ with momentum fraction $\xi_a$.  The Wilson coefficients
$\cI_{ik}(t_a, z_a, \mu_B)$ describe the collinear virtual and real ISR emitted
by this parton at the beam scale $\mu_B^2\simeq t_a \simeq \tau_B Q^2$. The real
ISR causes the formation of a jet prior to the hard collision
which is observed as radiation centered around the beam axis.
The PDFs in \eq{B_fact} are evaluated at the beam scale $\mu_B$, because
the measurement of $\tau_B$ introduces sensitivity to
the virtualities $t \simeq \tau_B Q^2$ of the colliding hard partons, giving
large logarithms $\ln(\mu_B^2/t)$.

For small $\tau_B$, we have
\begin{equation}
\tau_B = \frac{t_a}{Q^2} + \frac{t_b}{Q^2} + \frac{k_B}{Q} + \ord{\tau_B^2}
\,,\end{equation}
where the last term is the contribution from soft radiation at the scale
$\mu_S \simeq k_B \simeq \tau_B Q$ and is described by the soft function
$S_B(k_B, \mu_S)$ in \eq{DYbeamrun}. The collinear and soft contributions are
not separately measurable, which leads to the convolution of $S_B$, $B_i$, and $B_j$
in \eq{DYbeamrun}.  $S_B(k_B, \mu_S)$ includes the effects of
hadronization and soft radiation in the underlying event.  For $\lqcd \ll
\mu_S$, it is perturbatively calculable with power corrections of
$\ord{\lqcd/\mu_S}$.

The large logarithms $\alpha_s^n \ln^m\tau_B$, with $m \leq 2n$, are summed in
\eq{DYbeamrun} as follows. The hard, beam, and soft functions are each evaluated
at their natural scale $\abs{\mu_H} = Q$, $\mu_B = \sqrt{\tau_B} Q$, and $\mu_S =
\tau_B Q$, respectively, where they contain no large logarithms and can be computed in
fixed-order perturbation theory. They are then evolved to an arbitrary common
scale $\mu$ by the evolution kernels $U_H$, $U_B^{i,j}$, and $U_S$, and this
sums logarithms of the three scale ratios $\mu/\mu_H$, $\mu/\mu_B$, and $\mu/\mu_S$,
respectively. The combination of the different evolution kernels in
\eq{DYbeamrun} is $\mu$-independent and sums the logarithms of $\tau_B$. The hard
function for Drell-Yan production is a timelike form factor and for $\mu_H\simeq Q$
contains large $\pi^2$ terms from $\ln^2(-\img Q/\mu_H)$. We sum these
$\pi^2$ terms by taking $\mu_H=-\img Q$~\cite{pi2}.
We estimate perturbative uncertainties by varying $\mu_H$, $\mu_B$, and $\mu_S$
about the above values.  The complete summation at NNLL requires the
NLO expressions for $H_{ij}$, $\cI_{qq}$, $\cI_{qg}$, and $S_B$, as
well as the NNLL expressions for $U_H$, $U_B^{i,j}$, and $U_S$. See
Ref.~\cite{Stewart:2010qs} for a discussion and references of the required
anomalous dimensions and fixed-order computations.

\paragraph*{Results at NNLL.}

\begin{figure}[t!]
\includegraphics[width=\columnwidth]{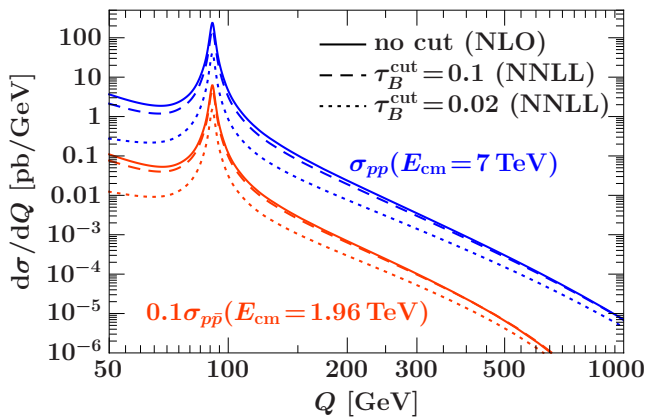}%
\vspace{-0.5ex}
\caption{
  Drell-Yan cross section $\df\sigma/\df Q$ with cuts $\tau_B \leq 0.1$ (dashed lines) and $\tau_B \leq 0.02$ (dotted lines) at NNLL. The solid lines show the total NLO cross section without a cut. For better distinction, the Tevatron cross section is multiplied by $0.1$.}
\label{fig:DYsigQ}
\end{figure}

\begin{figure}[t!]
\includegraphics[width=\columnwidth]{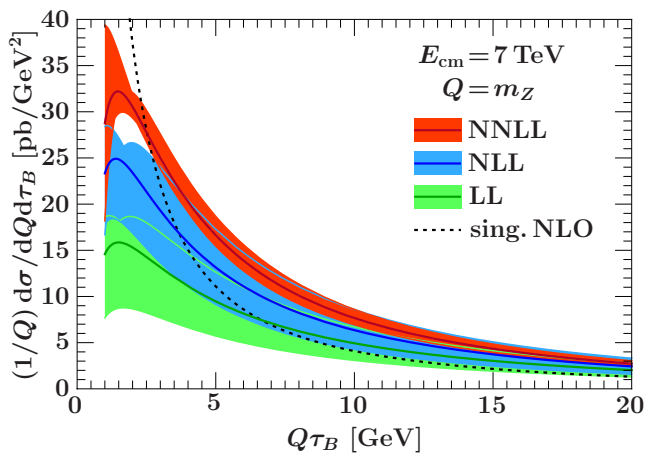}%
\vspace{-0.5ex}
\caption{
  Cross section differential in $\tau_B$ at $Q = m_Z$
  for the LHC with $\Ecm = 7\TeV$. Shown are the LL, NLL, and NNLL results, where the bands indicate the
  perturbative uncertainties as explained in the text. For comparison, the dotted line shows the
  singular NLO result with no resummation.}
\label{fig:DYdsdTauB}
\end{figure}

\begin{figure*}[t!]
\includegraphics[width=\columnwidth]{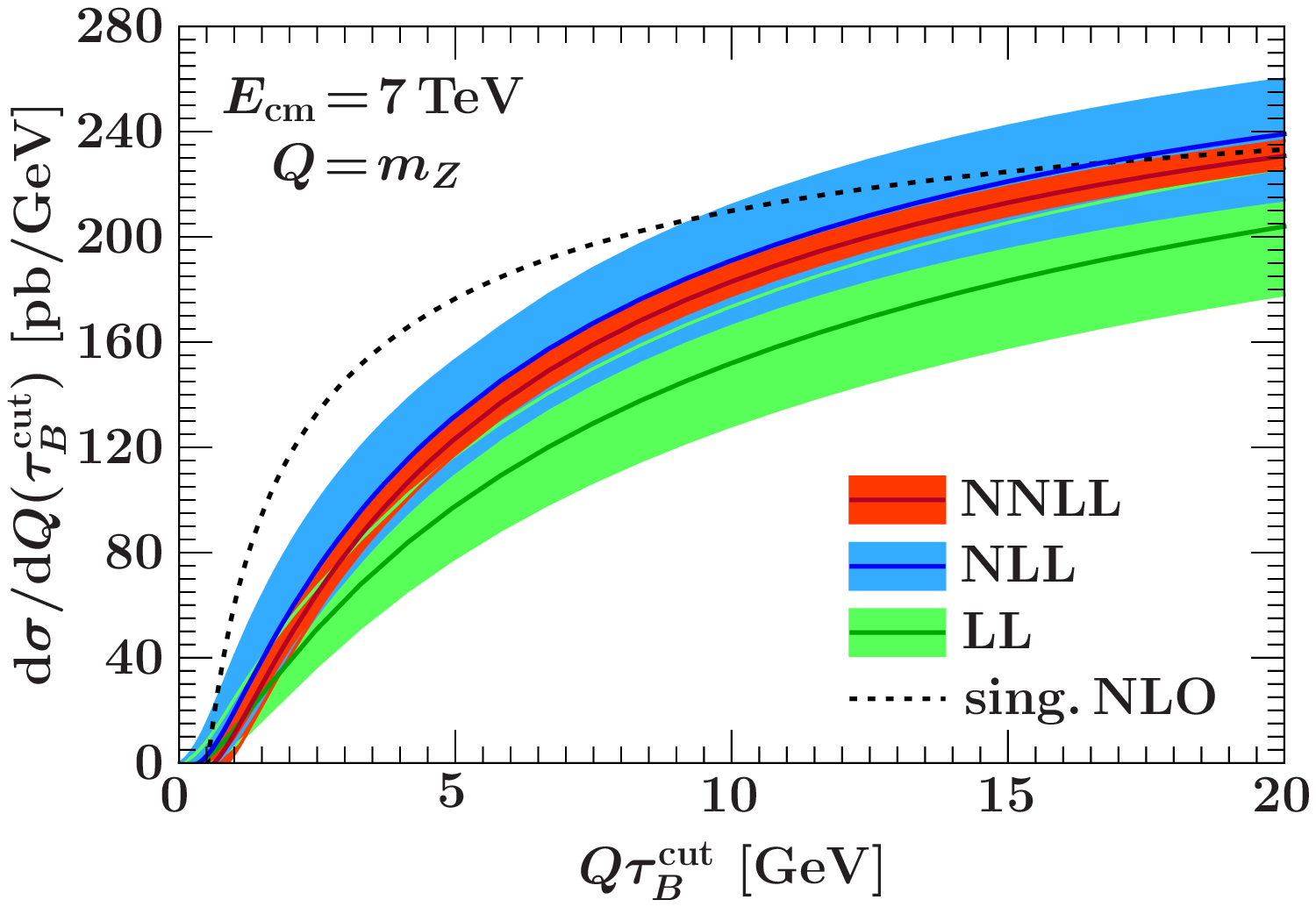}%
\hfill%
\includegraphics[width=\columnwidth]{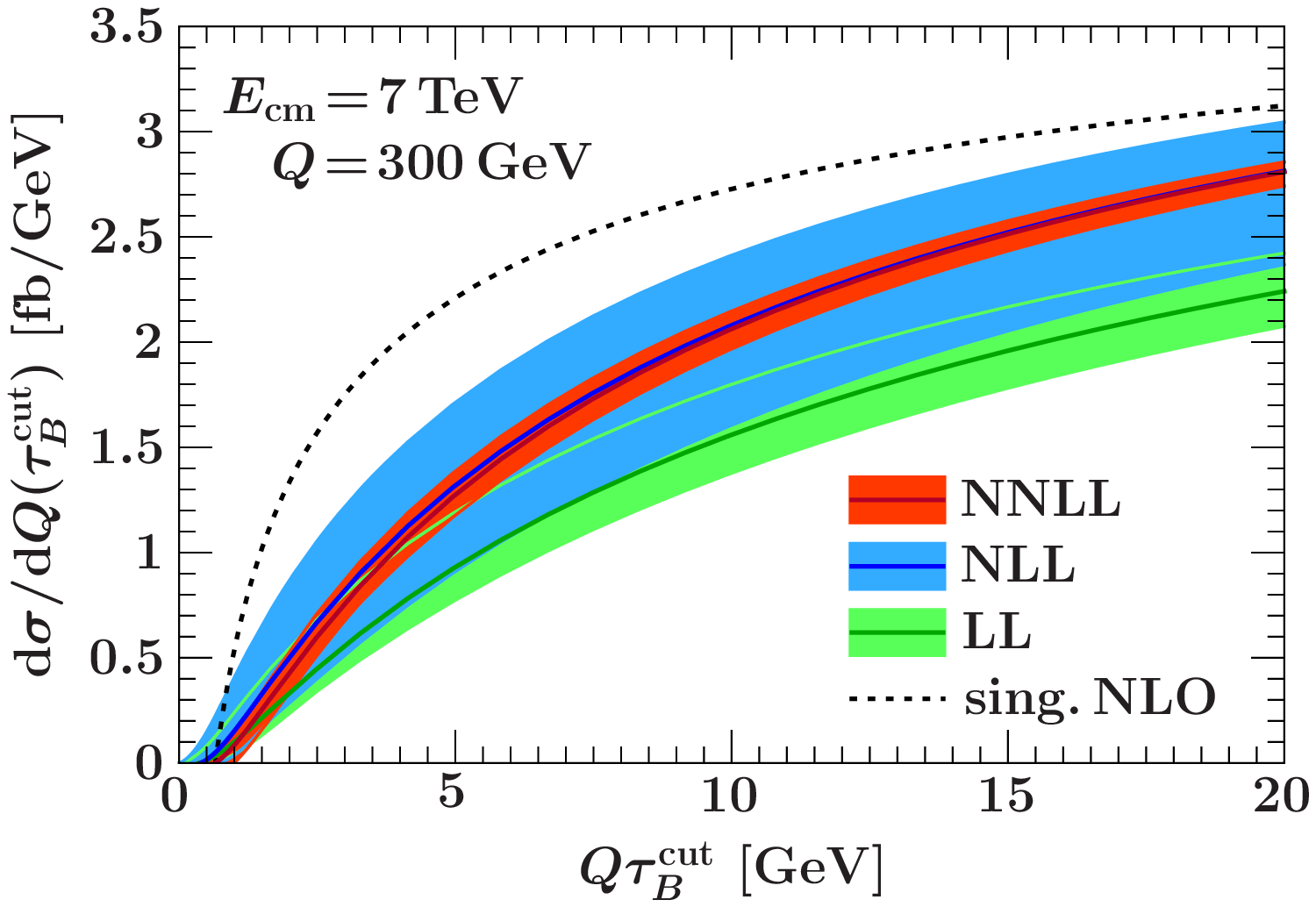}%
\vspace{-0.5ex}
\caption{
  Integrated cross section with a cut $\tau_B \leq \tau_B^\cut$ as a function of $\tau_B^\cut$ at $Q=m_Z$ (left panel) and $Q = 300\GeV$ (right panel) for the LHC. The curves have the same meaning as in \fig{DYdsdTauB}.}
\label{fig:DYsigTauBcut}
\end{figure*}

\begin{figure*}[t!]
\includegraphics[width=\columnwidth]{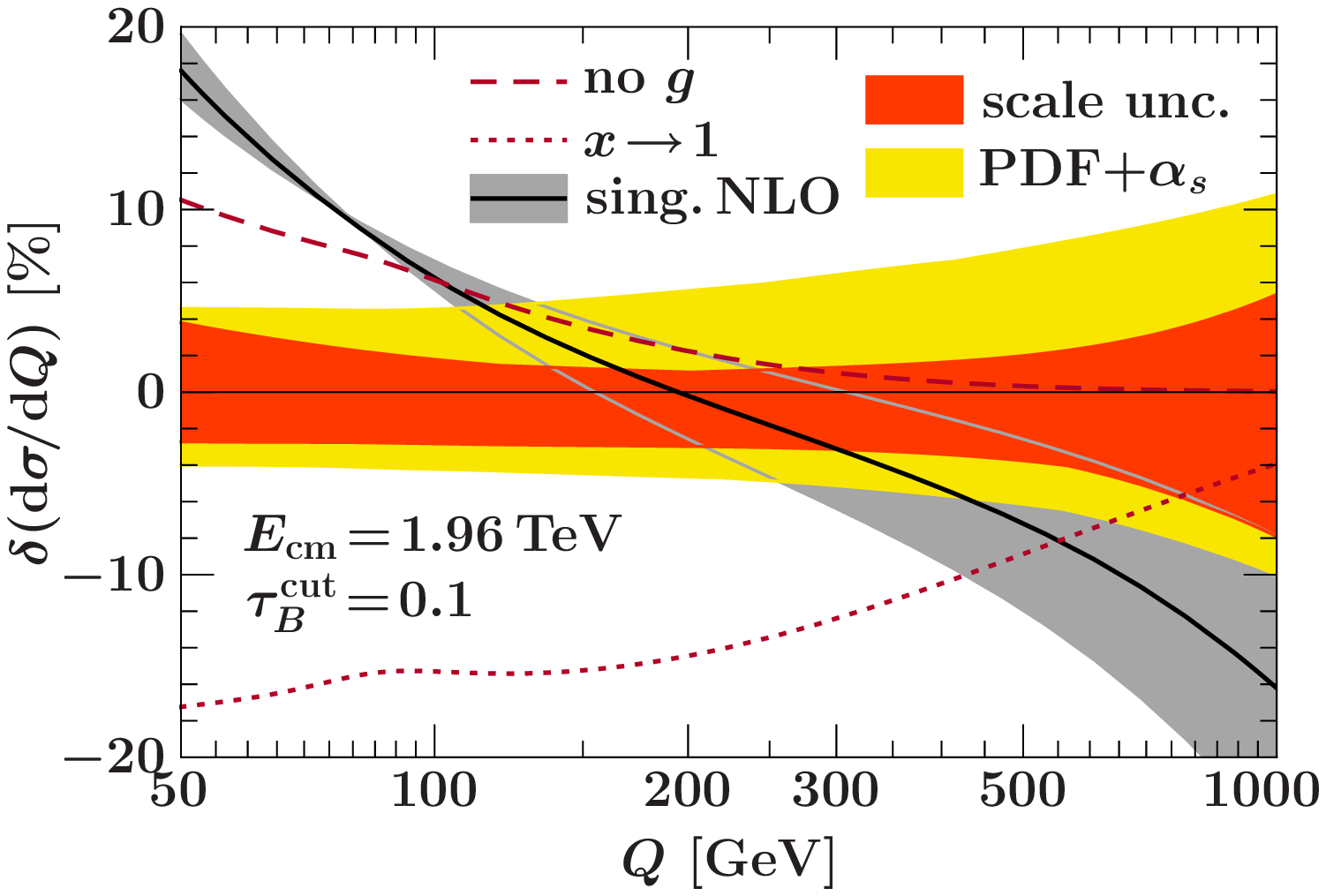}%
\hfill%
\includegraphics[width=\columnwidth]{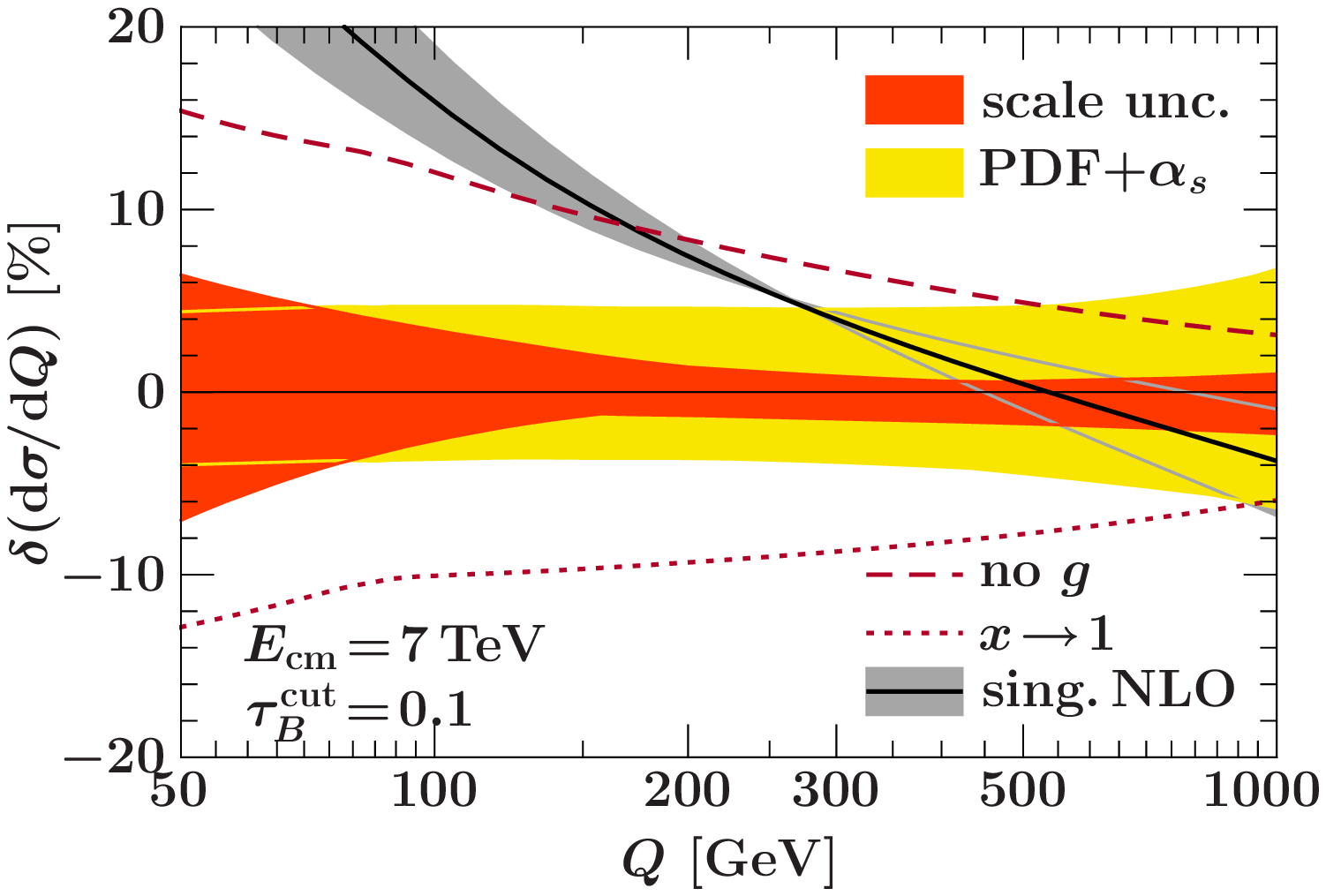}%
\vspace{-0.5ex}
\caption{
  Percent difference relative to the central NNLL result,
  $\delta(\df\sigma/\df Q) = (\df\sigma/\df Q)/(\df\sigma^\mathrm{NNLL}/\df Q) - 1$,
  at the Tevatron (left panel) and the LHC (right panel).
  Here we always take $\tau_B^\cut = 0.1$ in both the numerator and denominator.}
\label{fig:DYsigQ_unc}
\end{figure*}

In our numerical results we use the MSTW2008 NLO PDFs~\cite{MSTW08} with their
$\alpha_s(m_Z) = 0.1208$. We also integrate over $Y$ in \eq{DYbeamrun}.
In \fig{DYsigQ}, we show the Drell-Yan cross section $\df\sigma/\df Q$ with no
cut at NLO and with cuts $\tau_B \leq \{0.1, 0.02\}$ at NNLL, for the LHC with
$\Ecm = 7\TeV$ and for the Tevatron. The $Z$ resonance is visible at $Q = m_Z$.

The cut $\tau_B\leq 0.1$ reduces the cross section only by a factor of
around $1.3$ above the $Z$ peak (or $5$--$1.5$ for $\tau_B \leq 0.02$), showing
that most of the cross section comes from small $\tau_B$.
The cross section differential in $\tau_B$ at fixed $Q =
m_Z$ is shown in \fig{DYdsdTauB}, where we can see explicitly that the cross
section is dominated by small $\tau_B$. To see the effect of the higher-order
resummation we plot the LL, NLL, and NNLL results. The importance of resummation
is illustrated by comparing them to the singular part of the fixed NLO result
(dashed line), which is obtained from our NNLL result by setting $\mu_H = \mu_B
= \mu_S = Q$.  (The full NLO result contains additional nonsingular terms that
are not numerically relevant at small $\tau_B$.)  Results are not plotted below
$Q\tau_B = \mu_S \leq 1\GeV$, where the soft function becomes nonperturbative
and we expect large corrections of $\ord{\lqcd/\mu_S}$ to our purely perturbative results.
Correspondingly, the perturbative uncertainties get large here.  In
\fig{DYsigTauBcut} we show the cross section integrated up to $\tau_B \leq
\tau_B^\cut$ as a function of $Q\tau_B^\cut$ for $Q = m_Z$ and $Q = 300\GeV$. We
see again that the logarithms are important at small $\tau_B^\cut$ and need to
be resummed.

In Figs.~\ref{fig:DYdsdTauB} and \ref{fig:DYsigTauBcut}, the perturbative scale
uncertainties are given by bands from varying $\mu_H$, $\mu_B$, and $\mu_S$.
The independent variation of these three scales would overestimate the
uncertainty, since it does not take into account the parametric relation $\mu_B^2
\simeq \mu_S \mu_H$ and the hierarchy $\mu_S \ll \mu_B \ll \mu_H$.  On the other
hand, their simultaneous variation [case (a) in \eq{scales}] can produce
unnaturally small scale uncertainties.  Hence, the perturbative uncertainties in
all figures are the envelope of the separate scale variations
\begin{align} \label{eq:scales}
\text{(a)}&&
\mu_H &= -r \img Q
\,,\
\mu_B = r \sqrt{\tau_B} Q
\,,\
\mu_S = r \tau_B Q
\,,\\
\text{(b)}&&
\mu_H &= -\img Q
\,,\
\mu_B = r^{-(\ln \tau_B)/4} \sqrt{\tau_B}Q
\,,\
\mu_S = \tau_B Q
\,,\nn \\\nn
\text{(c)}&&
\mu_H &= -\img Q
\,,\
\mu_B = \sqrt{\tau_B} Q
\,,\
\mu_S = r^{-(\ln \tau_B)/4} \tau_B Q
\,,\end{align}
with $r = \{1/2, 2\}$, and $r = 1$ corresponding to the central-value curves.
The exponent of $r$ for cases (b) and (c) is chosen such that for $\tau_B =
e^{-4}$ the scales $\mu_B$ or $\mu_S$ vary by factors of $2$, with smaller
variations for increasing $\tau_B$ and no variation for $\tau_B \to 1$. In this
limit, there should only be a single scale $\abs{\mu_H} = \mu_B = \mu_S$, and thus the
only scale variation should be case (a). For the integrated cross section we replace
$\tau_B$ in \eq{scales} with $\tau_B^\cut$. In
both \figs{DYdsdTauB}{DYsigTauBcut}, we see good convergence of the
perturbative series and a substantial reduction in the perturbative
uncertainties at NNLL. The convergence is improved appreciably by the
summation of the $\pi^2$ terms.

In \fig{DYsigQ_unc}, we plot percent differences for several cross sections
relative to the NNLL result. All results are integrated up to $\tau_B^\cut =0.1$
and are plotted versus $Q$. The dark orange bands show the NNLL perturbative
uncertainties and the light yellow bands the 90\% C.L. PDF$+\alpha_s$
uncertainties using the procedure from Ref.~\cite{MSTW08}.  The dashed
line shows the NNLL result without the gluon contribution to the quark beam
function, $\cI_{qg}$ in \eq{B_fact}. The gluon contribution is significant at the
LHC and less prominent at the Tevatron, because the gluon PDF is more important
for $pp$ than $p\bar p$ collisions.  In the dotted line we further neglect all
terms in the quark contribution $\cI_{qq}$ that are subleading in the
threshold limit $x\to 1$.  Except for the Tevatron at large $Q$ the threshold
result is a poor approximation to the full result, being well outside the
perturbative uncertainties.  The dark band and
solid line show the NLO result with the perturbative uncertainties from varying
the common scale between $Q/2$ and $2Q$. Its difference from the resummed NNLL
result is generically large and not captured by the fixed-order perturbative
uncertainties, showing that the resummation is important not only to get an
improved central value but also to obtain reliable perturbative uncertainties.

Beam thrust in Drell-Yan production provides an experimentally and theoretically clean
measure of ISR in $q\bar{q} \to \ell^+\ell^-$, similar to how thrust measures
final-state radiation in $e^+e^-\to q\bar{q}$. The experimental measurement of
beam thrust will contribute very valuable information to our understanding of
ISR at hadron colliders and could be used to test and tune the initial-state
parton shower and underlying event models in Monte Carlo programs.
Restricting beam thrust $\tau_B \ll 1$ implements a theoretically well-controlled
jet veto, which has important applications in other processes, for example,
Higgs production~\cite{Berger:2010xi}. The measurement of beam thrust in Drell-Yan provides a
clean environment to test the application of beam thrust as a central jet veto.


This work was supported by the Office of Nuclear Physics of the U.S.\ Department
of Energy, under Grant No. DE-FG02-94ER40818.


\end{document}